\documentclass[showpacs,preprintnumbers,amsmath,amssymb]{revtex4}

\usepackage{graphicx}
\usepackage{dcolumn}
\usepackage{bm}
\usepackage{epsf}

\renewcommand{\bar}[1]{\overline{#1}}

\usepackage{amssymb}

\renewcommand{\bar}[1]{\overline{#1}}

\usepackage{indentfirst}
\usepackage{psfig,color}
\usepackage{epsfig}
\usepackage{epsf}
\usepackage{graphicx}
\providecommand{\Journal}[4] {#1 {\bf #2}, #3 (#4)}
\providecommand{\MPLA}{Mod. Phys. Lett. A} %
\providecommand{\NPA}{Nucl. Phys. A } %
\providecommand{\NPB}{Nucl. Phys. B } %
\providecommand{\PL}{Phys. Lett. } %
\providecommand{\PLB}{Phys. Lett. B } %
\providecommand{\PRL}{Phys. Rev. Lett. } %
\providecommand{\PRC}{Phys. Rev. C } %
\providecommand{\PRD}{Phys. Rev. D } %
\providecommand{\AP}{Annals Phys. } %
\providecommand{\PA}{Physica A } %

\begin{document}


\title{Radiative Decays of Decuplet Baryons, $\Lambda(1405)$ and $\Lambda
(1520)$ Hyperons}

\author{Lang Yu, Xiao-Lin Chen}
\affiliation{Department of Physics, Peking University, Beijing
100871, China}
\author{Wei-Zhen Deng}\email{dwz@th.phy.pku.edu.cn}
\affiliation{Department of Physics, Peking University, Beijing
 100871, China}
\author{Shi-Lin Zhu}\email{zhusl@th.phy.pku.edu.cn}
\affiliation{Department of Physics, Peking University, Beijing
100871, China}

\begin{abstract}

The chiral quark model gives a reasonably good description of many
low-energy observables by incorporating the effective degrees
carried by the constituent quarks and Goldstone bosons. We
calculate the decuplet to octet transition magnetic moments and
the decay widths of several excited hyperons using this model. The
various radiative decay widths from the chiral quark roughly agree
with experimental data including the recent JLAB measurement.

\end{abstract}





\vfill



\vfill





\vfill 



\maketitle

\section{INTRODUCTION}
\label{sec-intro}

The radiative decays of baryons contribute enormously to our
understanding of the underlying structure of baryons. The
non-relativistic quark model (NRQM) of Isgur and Karl
\cite{QM1,QM2} has been successful in predicting the
electromagnetic properties of the ground states of baryons $N$ and
their resonances $\emph{N}^*$. However, it is unable to give a
very good description of radiative decays of all decuplet and
other low-lying excited-state hyperons. Therefore, several other
theoretical approaches have been proposed to calculate these
transitions besides (NRQM)\cite{QM3,QM4}, including the
relativized constituent quark model (RCQM) \cite{RCQM}, the MIT
bag model \cite{QM3}, the chiral bag model \cite{CBM}, the Skyrme
model \cite{SKM}, the soliton model \cite{SOM}, the algebraic
model \cite{AM}, the heavy baryon chiral perturbation theory
(HB$\chi$PT) \cite{HBPT}, the $1/N_c$ expansion of QCD \cite{NM}
and the lattice calculations \cite{LA}.

Recently, one JLAB experiment \cite{CLAS} reported some new
results of the radiative decays of the $\Sigma^0(1385)$ and
$\Lambda(1520)$, suggesting that mesonic effects may play an
important role in $\Sigma^0(1385)$ radiative transitions
\cite{Meson}. On the other hand, a series of interesting work
about the chiral quark model \cite{XQM,XQM1,XQM2,XQM3} indicates
that the constituent quarks and internal Goldstone bosons can
offer an adequate description of flavor and spin structure of
baryons in the low $Q^2\lesssim 1$ GeV$^2$ region.  Within the
same framework, the octet and decuplet magnetic moments,
$\Sigma\Lambda$ transition magnetic moments and the explanation of
the violation of the Coleman-Glashow sum rule are in remarkably
good agreement with experimental data \cite{MM1,MM2}. So it is
interesting to explore whether we can make reliable predictions of
other important observables using the chiral quark model.

In this paper we calculate radiative decays of decuplet to octet
and some excited hyperons within the chiral model incorporating
quark sea perturbatively generated by the valence quark's emission
of internal Goldstone bosons. We can discern the contributions
from sea quarks and pseudoscalar mesons through the results of
transition magnetic moments and decay widths.

In Section \ref{sec-model}, we give an essential review of the
chiral quark model and the mechanism for the quark sea generation.
In Section \ref{sec-helicity}, we present the formalism of the
helicity amplitudes for the baryon radiative decays. In Section
\ref{sec-mm}, we present several typical cases of the calculation
of decuplet to octet transition magnetic moments. Then we
calculate the radiative decay widths of several excited hyperons
in \ref{sec-hyperon}. The numerical results and conclusions are
presented in the final section.

\section{MODEL DESCRIPTION}
\label{sec-model}

The chiral quark model \cite{XQM,XQM1,XQM2,XQM3} is an effective
theory of non-perturbative QCD, which is based on the interaction
between constituent quarks and Goldstone bosons. The basic process
is the emission of an internal Goldstone boson by a valence quark
:
\begin{equation}
q_{+}\longrightarrow GB+q'_{-}\longrightarrow(q\bar{q'})+q'_{-},
\end{equation}
or,\begin{equation} q_{+}\longrightarrow
GB+q'_{+}\longrightarrow(q\bar{q'})+q'_{+},
\end{equation}
where the subscripts indicate the helicity of the quark, and $[GB,
q']$ is in the helicity-flipping state ($\langle l_z \rangle =+1$)
in the process (1), and in the helicity-nonflipping state
($\langle l_z \rangle =0$) in the process (2). The quark may
change its helicity and flavor content by emitting a pseudoscalar
meson, and $(q\bar{q'})+q'$ constitute the quark sea
\cite{XQM,XQM4,XQM5,XQM6,XQM7,XQM8}. Thus, we consider the valence
constituent quark and the generted quark sea as a CQ-system
\cite{XQM8}. Moreover, the probability for the fluctuation of the
$q\bar{q}$ pairs is small because of the heavy mass of quarks in
the $Q^2\lesssim 1$ GeV$^2$ range. In other words, the interaction
is perturbative \cite{XQM,XQM4,XQM5,XQM8}. Therefore, the
effective Lagrangian describing the interaction between
constituent quarks and internal Goldstone bosons can be expressed
as follows,
\begin{equation}
\mathcal{L}_I=-\textrm{g}_8\bar{q}i\gamma^5\Phi q,
\end{equation}
\begin{equation}
\Phi=\left(\begin{array}{ccc}\frac{1}{\sqrt{2}}\pi^0+\beta\frac{1}{\sqrt{6}}
\eta+\zeta\frac{1}{\sqrt{3}}\eta'& \pi^+&\alpha K^+
\\\pi^-&-\frac{1}{\sqrt{2}}\pi^0+\beta\frac{1}{\sqrt{6}}\eta+\zeta\frac{1}
{\sqrt{3}}\eta'&\alpha K^0
\\\alpha K^-&\alpha\bar{K^0}&-\beta\frac{2}{\sqrt{6}}\eta+\zeta\frac{1}{\sqrt
{3}}\eta'\end{array}\right),
\end{equation}
where $q=(u, d, s)$, $\textrm{g}_1$ and $\textrm{g}_8$ denote the
coupling constants for the singlet and octet Goldstone bosons
respectively, and $\zeta=\textrm{g}_1/\textrm{g}_8$. Besides,
$\alpha$ and $\beta$ are introduced by considering SU(3) symmetry
breaking due to $\emph{M}_s>\emph{M}_{u,
d}$\cite{XQM4,XQM5,XQM6,XQM7}, whereas $\zeta$ is introduced by
considering the axial U(1) symmetry breaking
\cite{XQM,XQM4,XQM5,XQM6,XQM7}. Then, the transition probability
for the process $q\longrightarrow GB+q'$ can be easily deduced.
For example, $P(u\rightarrow
d+\pi^+)=a(a=\mid\textrm{g}_8\mid^2)$, $P(u\rightarrow
s+K^+)=\alpha^2a$, $P(u\rightarrow u+\pi^0)=\frac{1}{2}a$,
$P(u\rightarrow u+\eta)=\frac{1}{6}\beta^2a$, $P(u\rightarrow
u+\eta')=\frac{1}{3}\zeta^2a$ etc. Furthermore, because the total
angular momentum space wave function of the $[GB, q']$ state is
\begin{equation}
|J=\frac{1}{2},J_{z}=\frac{1}{2}\rangle=
\sqrt{\frac{2}{3}}|L=1,L_{z}=1\rangle|S=\frac{1}{2},S_{z}=-\frac{1}{2}\rangle
-\sqrt{\frac{1}{3}}|L=1,L_{z}=0\rangle|S=\frac{1}{2},S_{z}=\frac{1}{2}\rangle,
\end{equation}
the transition probability ratio of the process (1) to the process
(2) is $2:1$. Therefore, for example, $P(u_{+}\rightarrow
d_{-}+\pi^+)=\frac{2}{3}a$, and $P(u_{+}\rightarrow
d_{+}+\pi^+)=\frac{1}{3}a$.

In order to obtain the spin-flavor structure of the baryon, we
write the number operator $\mathcal{N}(j)$\cite{XQM7,MM2},
\begin{equation}
 \mathcal{N}(j)=\hat{n}_{u_+}(j)u_++\hat{n}_{u_-}(j)u_-+\hat{n}_{d_+}(j)
d_++\hat{n}_{d_-}(j)d_-
 +\hat{n}_{s_+}(j)s_++\hat{n}_{s_-}(j)s_-,
\end{equation}
where$\mathcal{N}(j)$ operates only on the jth quark,
$\hat{n}_{q_\pm}$ corresponds to the number operator of the
$q_\pm$. Thus, the spin-flavor structure of a baryon can be
defined as
\begin{equation}
\hat{B}=\sum^3_{j=1}\langle B | \mathcal{N}(j) | B \rangle\;.
\end{equation}
Using the symmetry of the baryon wave function, we can simplify
the equation above,
\begin{equation}
\hat{B}=\sum^3_{j=1}\langle B | \mathcal{N}(j) | B
\rangle=3\langle B | \mathcal{N}(3) | B \rangle,
\end{equation}
where $| B \rangle$ is the baryon wave function. Taking proton for
example and making use of the baryon $SU(6)\otimes O(3)$ wave
function\cite{WF}, we get
\begin{eqnarray}
\hat{p}&=&\frac{5}{3}u_++\frac{1}{3}u_-+\frac{1}{3}d_++\frac{2}{3}d_-\cr
&=&\frac{4}{3}u_+-\frac{1}{3}d_+.
\end{eqnarray}
Throughout our calculation, we make use of $q_-=-q_+$.

Furthermore, by considering the effects of $q\longrightarrow
GB+q'$, we need modify the baryon's spin-flavor structure above.
What we do is to make a replacement of every valence quark $q_\pm$
in the equation $\hat{B}$ as follows\cite{XQM7,MM2},
\begin{eqnarray}
q_\pm&\longrightarrow&(1-\sum_{q'= u, d, s}P_{(q\rightarrow
q')})q_\pm+\frac{1}{3}\sum_{q'= u, d, s}P_{(q\rightarrow
q')}(2q'_\mp+q'_\pm)+\frac{2}{3}\sum_{q'= u, d, s}P_{(q\rightarrow
q')}(q'_{\langle l^{GB}_{q'_{\pm}}\rangle}+GB_{\langle
l^{q'_{\mp}}_{GB}\rangle})\cr &\longrightarrow&(1-\sum_{q'= u, d,
s}P_{(q\rightarrow q')})q_\pm+\frac{1}{3}\sum_{q'= u, d,
s}P_{(q\rightarrow q')}q'_\mp+\frac{2}{3}\sum_{q'= u, d,
s}P_{(q\rightarrow q')}(q'_{\langle
l^{GB}_{q'_{\pm}}\rangle}+GB_{\langle l^{q'_{\mp}}_{GB}\rangle})
\end{eqnarray}
where $P_{(q\rightarrow q')}$ denotes the transition probability
of the process $q\longrightarrow GB+q'$, $q'_{\langle
l^{GB}_{q'_{\pm}}\rangle}$ and $GB_{\langle
l^{q'_{\pm}}_{GB}\rangle}$ denote that the orbit angular momenta
of the $q'$ and GB  are $\langle l^{GB}_{q'_{\pm}}\rangle$ and
$\langle l^{q'_{\pm}}_{GB}\rangle$ respectively(We neglect the
item of $\langle l_{z}\rangle=0$ in the orbit angular momenta
part, it does no contribution to the magnetic moments). And
$\langle l^{GB}_{q'_{\pm}}\rangle = \frac{M_{GB}}{M_{q'}+M_{GB}}
\langle l_z \rangle$, $\langle
l^{q'_{\pm}}_{GB}\rangle=\frac{M_{q'}}{M_{q'}+M_{GB}} \langle l_z
\rangle$(when $q'_{-}$,$\langle l_z \rangle=+1$; when $q'_{+}$,
$\langle l_z \rangle=-1$). In the equation (10), the first term
still corresponds to the valence quark spin. The latter two terms
are both from the contribution of quark sea. The second term
corresponds to the sea quark spin and the last term corresponds to
the orbit angular momenta between the sea quark and the Goldstone
boson. We take proton for example,
\begin{equation}
\hat{p}=\frac{4}{3}\{[1-a(\frac{9+\beta^2+2\zeta^2}{6}+\alpha^2)]u_+
+\frac{1}{3}[a(\frac{3+\beta^2+2\zeta^2}{6})u_-+ad_-+a\alpha^2s_-]+\frac{2}{3}
[a(\frac{1}{2})(u_{\langle l^{\pi^0}_{u_-}\rangle}+\pi^0_{\langle
l^{u_-}_{\pi^0}\rangle})\nonumber
\end{equation}
\begin{equation}
+a(\frac{\beta^2}{6})(u_{\langle
l^{\eta}_{u_-}\rangle}+\eta_{\langle
l^{u_-}_{\eta}\rangle})+a(\frac{\zeta^2}{3})(u_{\langle
l^{\eta'}_{u_-}\rangle}+\eta_{\langle
l^{u_-}_{\eta'}\rangle})+a(d_{\langle
l^{\pi^+}_{d_-}\rangle}+\pi^+_{\langle
l^{d_-}_{\pi^+}\rangle})+a\alpha^2(s_{\langle
l^{K^+}_{s_-}\rangle}+K^+_{\langle l^{s_-}_{K^+}\rangle})]\}
\nonumber
\end{equation}
\begin{equation}
-\frac{1}{3}\{[1-a(\frac{9+\beta^2+2\zeta^2}{6}+\alpha^2)]d_+
+\frac{1}{3}[a(\frac{3+\beta^2+2\zeta^2}{6})d_-+au_-+a\alpha^2s_-]+\frac{2}{3}
[a(\frac{1}{2})(d_{\langle l^{\pi^0}_{d_-}\rangle}+\pi^0_{\langle
l^{d_-}_{\pi^0}\rangle})\nonumber
\end{equation}
\begin{equation}
+a(\frac{\beta^2}{6})(d_{\langle
l^{\eta}_{d_-}\rangle}+\eta_{\langle
l^{d_-}_{\eta}\rangle})+a(\frac{\zeta^2}{3})(d_{\langle
l^{\eta'}_{d_-}\rangle}+\eta_{\langle
l^{d_-}_{\eta'}\rangle})+a(u_{\langle
l^{\pi^-}_{u_-}\rangle}+\pi^-_{\langle
l^{u_-}_{\pi^-}\rangle})+a\alpha^2(s_{\langle
l^{K^0}_{s_-}\rangle}+K^0_{\langle l^{s_-}_{K^0}\rangle})]\}
\end{equation}

In the next section, we need calculate the spin polarizations of
the  quark, following Refs.\cite{XQM,XQM5,XQM7,MM1,MM2}, which are
defined as
\begin{equation}
\Delta q=n_{q_+}-n_{q_-},
\end{equation}
where $n_{q_+}$ and $n_{q_-}$ can be calculated from the equation
$\hat{B}$. In this way, the expressions for proton can be obtained
as follows,
\begin{equation}
\Delta u^{val}=\frac{4}{3}
\end{equation}
\begin{equation}
\Delta d^{val}=-\frac{1}{3}
\end{equation}
\begin{equation}
\Delta
u^{sea}=-a(\frac{57+8\beta^2+16\zeta^2}{27}+\frac{4}{3}\alpha^2),
\end{equation}
\begin{equation}
\Delta
d^{sea}=a(\frac{6+4\beta^2+8\zeta^2}{54}+\frac{1}{3}\alpha^2),
\end{equation}
\begin{equation}
\Delta s^{sea}=-\frac{a\alpha^2}{3},
\end{equation}
Similarly, we can define $\Delta q^{orbit}_{GB}$ and $\Delta
GB^{orbit}_{q}$ as
\begin{equation}
\Delta q^{orbit}_{GB}=n_{q'_{\langle
l^{GB}_{q'_{-}}\rangle}}-n_{q'_{\langle l^{GB}_{q'_{+}}\rangle}},
\end{equation}
\begin{equation}
\Delta GB^{orbit}_{q}=n_{GB_{\langle
l^{q'_{-}}_{GB}\rangle}}-n_{GB_{\langle l^{q'_{+}}_{GB}\rangle}},
\end{equation}
Thus, we get the expressions for proton,
\begin{equation}
\Delta u^{orbit}_{\pi^0}=\frac{4}{9}a, \;\;\Delta \pi
^{0orbit}_{u}=\frac{4}{9}a
\end{equation}
\begin{equation}
\Delta u^{orbit}_{\eta}=\frac{4\beta^2}{27}a, \;\;\Delta
\eta^{orbit}_{u}=\frac{4\beta^2}{27}a
\end{equation}
\begin{equation}
\Delta u^{orbit}_{\eta'}=\frac{8\zeta^2}{27}a,\;\;\Delta
\eta'^{orbit}_{u}=\frac{8\zeta^2}{27}a
\end{equation}
\begin{equation}
\Delta d^{orbit}_{\pi^+}=\frac{8}{9}a,\;\;\Delta \pi^
{+orbit}_{d}=\frac{8}{9}a
\end{equation}
\begin{equation}
\Delta s^{orbit}_{K^+}=\frac{8\alpha^2}{9}a,\;\;\Delta
K^{+orbit}_{s}=\frac{8\alpha^2}{9}a
\end{equation}
\begin{equation}
\Delta d^{orbit}_{\pi^0}=-\frac{1}{9}a,\;\;\Delta
\pi^{0orbit}_{d}=-\frac{1}{9}a
\end{equation}
\begin{equation}
\Delta d^{orbit}_{\eta}=-\frac{\beta^2}{27}a,\;\;\Delta
\eta^{orbit}_{d}=-\frac{\beta^2}{27}a
\end{equation}
\begin{equation}
\Delta d^{orbit}_{\eta'}=-\frac{2\zeta^2}{27}a,\;\;\Delta
\eta'^{orbit}_{d}=-\frac{2\zeta^2}{27}a
\end{equation}
\begin{equation}
\Delta u^{orbit}_{\pi^-}=-\frac{2}{9}a,\;\;\Delta
\pi^{-orbit}_{u}=-\frac{2}{9}a
\end{equation}
\begin{equation}
\Delta s^{orbit}_{K^0}=-\frac{2\alpha^2}{9}a,\;\;\Delta
K^{0orbit}_{s}=-\frac{2\alpha^2}{9}a
\end{equation}

\section{Helicity Amplitudes of Radiative Decays}
\label{sec-helicity}

In addition, we give a short review of the helicity amplitude in
order to calculate radiative decay widths \cite{RT,WF}. In case of
the process $B_i\longrightarrow B_f+\gamma$,
\begin{equation}
A_M=-e\sqrt{{2\pi}/{k}}\langle B_f, J_z=M-1|\varepsilon^* \cdot
\sum^3_{i=1}j_{em}(i)|B_i, J_z=M \rangle\;\;M= \frac{3}{2},
\frac{1}{2}
\end{equation}
where \textbf{k} and $\varepsilon$ are the momentum and
polarization vector of the photon, and $j_{em}(i)$ is the ith
quark current density. Without loss of generality, we take the
photon to be right-handed [$\varepsilon=-1/\sqrt{2}(1,i,0)$] and
expand $\varepsilon^* \cdot \sum^3_{i=1}j_{em}(i)$,
\begin{equation}
\varepsilon^*\cdot\sum^3_{i=1}j_{em}(i)=-\sqrt{2}\sum^3_{i=1}\frac{e}{2m_i}[e^
{-\textrm{i}kz(i)}q(i)][k\sigma_-(i)+(p_x(i)-\textrm{i}p_y(i))]\;.
\end{equation}
With the symmetry of the baryon wave function, we can simplify the
above equation,
\begin{eqnarray}
  \varepsilon^*\cdot\sum^3_{i=1}j_{em}(i)
 &=&-\sqrt{2}\sum^3_{i=1}\frac{e}{2m_i}[e^{-\textrm{i}kz(i)}q(i)][k\sigma_-
(i)+(p_x(i)-\textrm{i}p_y(i))] \cr
 &=&-\sqrt{2}\frac{3e}{2m_3}[e^{-\textrm{i}kz(3)}q(3)][k\sigma_-(3)+(p_x(3)-
\textrm{i}p_y(3))]
\end{eqnarray}
where the first term contributes to magnetic moments or
magnetic-dipole transitions, and the second term contributes to
electric-dipole transitions between $L=1$ orbit excitations and
the ground states. Thus, we can obtain the radiative widths in
term of $A_{\frac{1}{2}}$ and $A_{\frac{3}{2}}$,
\begin{equation}
\Gamma=\frac{k^2}{2\pi}\frac{1}{2J+1}\frac{m_f}{m_i}\{|A_\frac{3}{2}
|^2+|A_\frac{1}{2}|^2\}
\end{equation}

\section{Decuplet to octet transition magnetic moments}
\label{sec-mm}

In this section we calculate the decuplet to octet transition
magnetic moments. Making use of the equations (12), (18), (19),
we can express the magnetic moment of a given quark B as
\begin{equation}
\mu_B^{total}=\mu_B^{val}+\mu_B^{sea}+\mu_B^{orbit}
\end{equation}
where
\begin{equation}
\mu_B^{val}=\sum_{q=u,d,s}\Delta q^{val}\mu_q
\end{equation}
\begin{equation}
\mu_B^{sea}=\sum_{q=u,d,s}\Delta q^{sea}\mu_q
\end{equation}
\begin{equation}
\mu_B^{orbit}=\sum_{q=u,d,s}\Delta q^{orbit}_{GB}\mu_q\langle
l^{GB}_{q_{-}}\rangle+\sum_{GB}\Delta
GB^{orbit}_{q}\mu_{GB}\langle l^{q_{-}}_{GB}\rangle
\end{equation}

Similarly, we can get the transition magnetic moments of
$B_{10}\longrightarrow B_8+\gamma$ transitions. In the absence of
the conventional form factor, we can use the equations (34)-(37)
directly after calculating
$\widehat{B_{10}B_8}=\sum^3_{j=1}\langle B_8, J_z=\frac{1}{2} |
\mathcal{N}(j) | B_{10}, J_z=\frac{1}{2} \rangle$\cite{MM2}.
However, if we consider the form factor, $A_M$ contains the
integral like
$\langle\psi^s_{000}|e^{-\textrm{i}kz(3)}|\psi^s_{000}\rangle$
\cite{WF,RT} for the radiative decays between baryons with $L_i=0$
and $L_f=0$ if we define the baryon flavor-spin-space wave
function as $| B \rangle=\phi\chi\psi$. Therefore, we need make
slight modification of the $\hat{B}$ and get $\hat{B}(k)$,
\begin{eqnarray}
\widehat{B_{10}B_8}(k) &=&\sum^3_{j=1}\langle B_8, J_z=\frac{1}{2}
| \mathcal{N}(j)\cdot e^{-\textrm{i}kz(j)} | B_{10} ,
J_z=\frac{1}{2}\rangle =3\langle B_8, J_z=\frac{1}{2} |
\mathcal{N}(3)\cdot e^{-\textrm{i}kz(3)} | B_{10}, J_z=
\frac{1}{2} \rangle\cr
&=&\widehat{B_{10}B_8}\cdot\langle\psi^s_{000}|e^{-\textrm{i}kz(3)}|\psi^s_
{000}\rangle,
\end{eqnarray}
where \textbf{k} is the momentum of the photon. We add spatial
parts to the operator and thus need add the form factor to the
transition magnetic moments,
\begin{equation}
\mu_{B_{10}B_8}(k)=\mu_{B_{10}B_8}\cdot\langle\psi^s_{000}|e^{-\textrm{i}kz
(3)}|\psi^s_{000}\rangle
\end{equation}
Because $A_M$ contains contribution only from magnetic-dipole
transitions ($M1$) for the decuplet to octet transitions, we can
write $A_M$ in terms of $\mu_{B_{10}B_8}(k)$,
\begin{equation}
A_{\frac{3}{2}}=A_{\frac{3}{2}}^{M1}=-\sqrt{3{\pi}{k}}\cdot\mu_{B_{10}B_8}(k)
\end{equation}
\begin{equation}
A_{\frac{1}{2}}=A_{\frac{1}{2}}^{M1}=-\sqrt{{\pi}{k}}\cdot\mu_{B_{10}B_8}(k)
\end{equation}

Next, we list the detailed calculations of
$\Sigma^0(1385)\longrightarrow\Lambda(1116)+\gamma$ ,
\begin{eqnarray}
\widehat{\Sigma^{*,0}\Lambda}(k)
&=&\widehat{\Sigma^{*,0}\Lambda}\cdot\langle\psi^s_{000}|e^{-\textrm{i}kz(3)}
|\psi^s_{000}\rangle\cr &=&\frac{\sqrt{6}}{3}(u^+-d^+)\cdot
e^{-\frac{1}{6}k^2R^2}
\end{eqnarray}
considering  $q\longrightarrow GB+q'$,

\begin{eqnarray}
\widehat{\Sigma^{*,0}\Lambda}(k)
&=&\frac{\sqrt{6}}{3}\{[1-a(\frac{9+\beta^2+2\zeta^2}{6}+\alpha^2)]u_+
+\frac{1}{3}[a(\frac{3+\beta^2+2\zeta^2}{6})u_-+ad_-+a\alpha^2s_-]+\frac{2}{3}
[a(\frac{1}{2})(u_{\langle l^{\pi^0}_{u_-}\rangle}+\pi^0_{\langle
l^{u_-}_{\pi^0}\rangle})\cr &+&a(\frac{\beta^2}{6})(u_{\langle
l^{\eta}_{u_-}\rangle}+\eta_{\langle
l^{u_-}_{\eta}\rangle})+a(\frac{\zeta^2}{3})(u_{\langle
l^{\eta'}_{u_-}\rangle}+\eta_{\langle
l^{u_-}_{\eta'}\rangle})+a(d_{\langle
l^{\pi^+}_{d_-}\rangle}+\pi^+_{\langle
l^{d_-}_{\pi^+}\rangle})+a\alpha^2(s_{\langle
l^{K^+}_{s_-}\rangle}+K^+_{\langle l^{s_-}_{K^+}\rangle})] \cr
&-&[1-a(\frac{9+\beta^2+2\zeta^2}{6}+\alpha^2)]d_+
-\frac{1}{3}[a(\frac{3+\beta^2+2\zeta^2}{6})d_-+au_-+a\alpha^2s_-]-\frac{2}{3}
[a(\frac{1}{2})(d_{\langle l^{\pi^0}_{d_-}\rangle}+\pi^0_{\langle
l^{d_-}_{\pi^0}\rangle})\cr &+&a(\frac{\beta^2}{6})(d_{\langle
l^{\eta}_{d_-}\rangle}+\eta_{\langle
l^{d_-}_{\eta}\rangle})+a(\frac{\zeta^2}{3})(d_{\langle
l^{\eta'}_{d_-}\rangle}+\eta_{\langle
l^{d_-}_{\eta'}\rangle})+a(u_{\langle
l^{\pi^-}_{u_-}\rangle}+\pi^-_{\langle
l^{u_-}_{\pi^-}\rangle})+a\alpha^2(s_{\langle
l^{K^0}_{s_-}\rangle}+K^0_{\langle l^{s_-}_{K^0}\rangle})]\} \cr
&\cdot& e^{-\frac{1}{6}k^2R^2}
\end{eqnarray}

\begin{equation}
\mu_{\Sigma^{*,0}\Lambda}^{total}(k)=\mu_{\Sigma^{*,0}\Lambda}^{val}(k)
+\mu_{\Sigma^{*,0}\Lambda}^{sea}(k)+\mu_{\Sigma^{*,0}\Lambda}^{orbit}(k)
\end{equation}

\begin{eqnarray}
\mu_{\Sigma^{*,0}\Lambda}^{val}(k)
&=&\mu_{\Sigma^{*,0}\Lambda}^{val}\cdot e^{-\frac{1}{6}k^2R^2}\cr
&=&\frac{\sqrt{6}}{3}(\mu_u-\mu_d) \cdot e^{-\frac{1}{6}k^2R^2}
\end{eqnarray}

\begin{eqnarray}
\mu_{\Sigma^{*,0}\Lambda}^{sea}(k)
&=&\mu_{\Sigma^{*,0}\Lambda}^{sea}\cdot e^{-\frac{1}{6}k^2R^2}\cr
&=&-\frac{\sqrt{6}}{3}a(\frac{12+2\beta^2+4\zeta^2}{9}+\alpha^2)(\mu_u-\mu_d)\cdot
e^{-\frac{1}{6}k^2R^2}
\end{eqnarray}

\begin{eqnarray}
\mu_{\Sigma^{*,0}\Lambda}^{orbit}(k)
&=&\mu_{\Sigma^{*,0}\Lambda}^{orbit}\cdot
e^{-\frac{1}{6}k^2R^2}\cr
&=&\frac{2\sqrt{6}}{9}a[\frac{1}{2}(\mu_u{\langle
l^{\pi^0}_{u_-}\rangle}+\mu_{\pi^0}{\langle
l^{u_-}_{\pi^0}\rangle})+\frac{\beta^2}{6}(\mu_u{\langle
l^{\eta}_{u_-}\rangle}+\mu_{\eta}{\langle
l^{u_-}_{\eta}\rangle})+\frac{\zeta^2}{3}(\mu_u{\langle
l^{\eta'}_{u_-}\rangle}+\mu_{\eta}{\langle
l^{u_-}_{\eta'}\rangle})\cr &+&(\mu_d{\langle
l^{\pi^+}_{d_-}\rangle}+\mu_{\pi^+}{\langle
l^{d_-}_{\pi^+}\rangle})+\alpha^2(\mu_s{\langle
l^{K^+}_{s_-}\rangle}+\mu_{K^+}{\langle l^{s_-}_{K^+}\rangle})\cr
&-&\frac{1}{2}(\mu_d{\langle
l^{\pi^0}_{d_-}\rangle}+\mu_{\pi^0}{\langle
l^{d_-}_{\pi^0}\rangle})-\frac{\beta^2}{6}(\mu_d{\langle
l^{\eta}_{d_+}\rangle}+\mu_{\eta}{\langle
l^{d_+}_{\eta}\rangle})-\frac{\zeta^2}{3}(\mu_d{\langle
l^{\eta'}_{d_+}\rangle}+\mu_{\eta}{\langle
l^{d_+}_{\eta'}\rangle})\cr &-&(\mu_u{\langle
l^{\pi^-}_{u_+}\rangle}+\mu_{\pi^-}{\langle
l^{u_+}_{\pi^-}\rangle})-\alpha^2(\mu_s{\langle
l^{K^0}_{s_+}\rangle}+\mu_{K^0}{\langle l^{s_+}_{K^0}\rangle})]
e^{-\frac{1}{6}k^2R^2}
\end{eqnarray}
So,
\begin{equation}
A_{\frac{3}{2}}^{total}(\Sigma^{*,0}\rightarrow\Lambda+\gamma)
=A_{\frac{3}{2}}^{M1,val}(\Sigma^{*,0}\rightarrow\Lambda+\gamma)
+A_{\frac{3}{2}}^{M1,sea}(\Sigma^{*,0}\rightarrow\Lambda+\gamma)
+A_{\frac{3}{2}}^{M1,orbit}(\Sigma^{*,0}\rightarrow\Lambda+\gamma)
\end{equation}
where,
\begin{equation}
A_{\frac{3}{2}}^{M1,val}(\Sigma^{*,0}\rightarrow\Lambda+\gamma)
=-\sqrt{3{\pi}{k}}\cdot\mu_{\Sigma^{*,0}\Lambda}^{val}(k)
\end{equation}
\begin{equation}
A_{\frac{3}{2}}^{M1,sea}(\Sigma^{*,0}\rightarrow\Lambda+\gamma)
=-\sqrt{3{\pi}{k}}\cdot\mu_{\Sigma^{*,0}\Lambda}^{sea}(k)
\end{equation}
\begin{equation}
A_{\frac{3}{2}}^{M1,orbit}(\Sigma^{*,0}\rightarrow\Lambda+\gamma)
=-\sqrt{3{\pi}{k}}\cdot\mu_{\Sigma^{*,0}\Lambda}^{orbit}(k)
\end{equation}
\begin{equation}
A_{\frac{1}{2}}^{total}(\Sigma^{*,0}\rightarrow\Lambda+\gamma)
=A_{\frac{1}{2}}^{M1,val}(\Sigma^{*,0}\rightarrow\Lambda+\gamma)
+A_{\frac{1}{2}}^{M1,sea}(\Sigma^{*,0}\rightarrow\Lambda+\gamma)
+A_{\frac{1}{2}}^{M1,orbit}(\Sigma^{*,0}\rightarrow\Lambda+\gamma)
\end{equation}
where,
\begin{equation}
A_{\frac{1}{2}}^{M1,val}(\Sigma^{*,0}\rightarrow\Lambda+\gamma)
=-\sqrt{{\pi}{k}}\cdot\mu_{\Sigma^{*,0}\Lambda}^{val}(k)
\end{equation}
\begin{equation}
A_{\frac{1}{2}}^{M1,sea}(\Sigma^{*,0}\rightarrow\Lambda+\gamma)
=-\sqrt{{\pi}{k}}\cdot\mu_{\Sigma^{*,0}\Lambda}^{sea}(k)
\end{equation}
\begin{equation}
A_{\frac{1}{2}}^{M1,orbit}(\Sigma^{*,0}\rightarrow\Lambda+\gamma)
=-\sqrt{{\pi}{k}}\cdot\mu_{\Sigma^{*,0}\Lambda}^{orbit}(k)
\end{equation}
besides,
\begin{equation}
A_{\frac{3}{2}}^{total}(\Sigma^{*,0}\rightarrow\Lambda+\gamma)
=-\sqrt{3{\pi}{k}}\cdot\mu_{\Sigma^{*,0}\Lambda}^{total}(k)
\end{equation}
\begin{equation}
A_{\frac{1}{2}}^{total}(\Sigma^{*,0}\rightarrow\Lambda+\gamma)
=-\sqrt{{\pi}{k}}\cdot\mu_{\Sigma^{*,0}\Lambda}^{total}(k)
\end{equation}
\begin{equation}
\Gamma(\Sigma^{*,0}\rightarrow\Lambda+\gamma)
=\frac{k^3}{2}\frac{m_{\Lambda}}{m_{\Sigma^{*,0}}}
[\mu_{\Sigma^{*,0}\Lambda}^{total}(k)]^2
\end{equation}

\section{Radiative decays of $\Lambda(1405)$ and $\Lambda
(1520)$} \label{sec-hyperon}

Both the magnetic-dipole transitions and electric-dipole
transitions contribute to the radiative decays of low-lying
excited ($L=1$) hyperons.
\begin{equation}
A_{M}=A_{M}^{M1}+A_{M}^{E1},\;\;M=\frac{1}{2} \;\;or
\;\;\frac{3}{2}
\end{equation}
We have to take the calculations of both
$\widehat{B_{i}B_f}_{(M)}(k)$(corresponding to $A_{M}^{M1}$) and
$\widehat{B_{i}B_f}^*_{(M)}(k)$(corresponding to $A_{M}^{E1}$),
\begin{eqnarray}
\widehat{B_{i}B_f}_{(M)}(k) &=&\sum^3_{j=1}\langle B_f, J_z=M |
\mathcal{N}(j)\cdot e^{-\textrm{i}kz(j)} | B_{i} , J_z=M\rangle\cr
&=&3\langle B_f, J_z=M | \mathcal{N}(3)\cdot e^{-\textrm{i}kz(3)}
| B_{i}, J_z= M \rangle,\;\;M=-\frac{1}{2} \;\;or \;\;\frac{1}{2}
\end{eqnarray}
and
\begin{eqnarray}
\widehat{B_{i}B_f}^*_{(M)}(k) &=&\frac{1}{k}\sum^3_{j=1}\langle
B_f, J_z=M-1 | \mathcal{N}^*(j)\cdot
e^{-\textrm{i}kz(j)}(p_x(i)-\textrm{i}p_y(i)) | B_{i} ,
J_z=M\rangle\cr &=&\frac{3}{k}\langle B_f, J_z=M-1 |
\mathcal{N}^*(3)\cdot e^{-\textrm{i}kz(3)}
(p_x(3)-\textrm{i}p_y(3))| B_{i}, J_z= M \rangle,\;\;M=\frac{1}{2}
\;\;or \;\;\frac{3}{2}
\end{eqnarray}
where $\mathcal{N}^*(j)$ is defined as
\begin{equation}
 \mathcal{N}^*(j)=\hat{n}_{u}(j)u+\hat{n}_{d}(j)d+\hat{n}_{s}(j)s.
\end{equation}
$\widehat{B_{i}B_f}(k)$ corresponds to the magnetic-dipole
transitions, which is discussed in Sec. A, while
$\widehat{B_{i}B_f}^*(k)$ corresponds to the electric-dipole
transitions, which will be explained below.

From the equation (31), the magnetic term flips the spin of the
quark and transforms as $\sigma_-$\cite{PE}. We must ensure that
the CQ-system remains to be a spin $\frac{1}{2}$ entity as a
single valence quark \cite{XQM8} when considering the effects of
$q\longrightarrow GB+q'$. Similarly, the electric term flips the
$L_z$ of the quark and transforms as $L_-$. So we calculate
spin-flavor structure $\widehat{B_{i}B_f}^*(k)$ and must ensure
that the CQ-system remains the same $L_z$ as a single valence
quark. Besides, because the electric term has no spin operators,
$\mathcal{N}^*(j)$ does not need spin subscripts.

Therefore, the $\widehat{B_{i}B_f}^*(k)$ can be calculated as
follows: first we calculate the $\widehat{B_{i}B_f}^*(k)$ without
considering $q\longrightarrow GB+q'$, then we replace every
valence quark as
\begin{equation}
q\longrightarrow(1-\sum_{q'= u, d, s}P_{(q\rightarrow
q')})q+\sum_{q'= u, d, s}P_{(q_\rightarrow q')}(q'_{\langle
l^{GB}_{q'_{}}\rangle}+GB_{\langle l^{q'_{}}_{GB}\rangle})
\end{equation}
where $\langle l^{GB}_{q'}\rangle = \frac{M_{GB}}{M_{q'}+M_{GB}}$,
and  $\langle l^{q'_{}}_{GB}\rangle=\frac{M_{q'}}{M_{q'}+M_{GB}}$.

We take $\Lambda(1520)\rightarrow\Lambda(1116)+\gamma$ for
example.
\begin{equation}
| \Lambda(1520),J_z=\frac{3}{2}
\rangle=\frac{1}{\sqrt{2}}(\phi^a\chi^{\lambda}_{\frac{1}{2}}\psi^{\rho}_{111}
-\phi^a\chi^{\rho}_{\frac{1}{2}}\psi^{\lambda}_{111}),
\end{equation}
\begin{equation}
| \Lambda(1520),J_z=\frac{1}{2}
\rangle=\frac{1}{\sqrt{6}}(\phi^a\chi^{\lambda}_{-\frac{1}{2}}\psi^{\rho}_
{111} -\phi^a\chi^{\rho}_{-\frac{1}{2}}\psi^{\lambda}_{111})
+\frac{1}{\sqrt{3}}(\phi^a\chi^{\lambda}_{\frac{1}{2}}\psi^{\rho}_{110}
-\phi^a\chi^{\rho}_{\frac{1}{2}}\psi^{\lambda}_{110}),
\end{equation}
\begin{equation}
| \Lambda(1116),J_z=\frac{1}{2}
\rangle=\frac{1}{\sqrt{2}}(\phi^{\rho}_{\Lambda}\chi^{\rho}_{\frac{1}{2}}
+\phi^{\lambda}_{\Lambda}\chi^{\lambda}_{\frac{1}{2}})\psi^{s}_{000}\;.
\end{equation}
With these wave functions, we obtain
\begin{eqnarray}
\widehat{\Lambda_{1520}\Lambda_{1116}}_{(-\frac{1}{2})}(k)=\textrm{i}\frac{1}
{6}(u_++d_+-2s_+)\cdot kR\cdot e^{-\frac{1}{6}k^2R^2},
\end{eqnarray}
in which
$\langle\psi^s_{000}|e^{-\textrm{i}kz(3)}|\psi^{\lambda}_{110}\rangle=\textrm
{i}\frac{\sqrt{3}}{3}\cdot kR\cdot e^{-\frac{1}{6}k^2R^2}$.
\begin{eqnarray}
\widehat{\Lambda_{1520}\Lambda_{1116}}_{(\frac{3}{2})}^*(k)=-\textrm{i}\frac
{\sqrt{3}}{6}(u+d-2s)\cdot \frac{1}{kR}\cdot
e^{-\frac{1}{6}k^2R^2},
\end{eqnarray}
\begin{eqnarray}
\widehat{\Lambda_{1520}\Lambda_{1116}}_{(\frac{1}{2})}^*(k)=-\textrm{i}\frac
{1}{6}(u+d-2s)\cdot \frac{1}{kR}\cdot e^{-\frac{1}{6}k^2R^2},
\end{eqnarray}
in which
$\langle\psi^s_{000}|e^{-\textrm{i}kz(3)}(p_x(3)-\textrm{i}p_y(3))|\psi^
{\lambda}_{111}\rangle=-\textrm{i}\frac{\sqrt{6}}{3}\cdot
\frac{1}{R}\cdot e^{-\frac{1}{6}k^2R^2}$.

Then, considering  $q\longrightarrow GB+q'$, we give a
modification of the equations (67)-(69).
\begin{eqnarray}
&&\widehat{\Lambda_{1520}\Lambda_{1116}}_{(-\frac{1}{2})}(k)\cr
&=&\{[1-a(\frac{9+\beta^2+2\zeta^2}{6}+\alpha^2)]u_+
+\frac{1}{3}[a(\frac{3+\beta^2+2\zeta^2}{6})u_-+ad_-+a\alpha^2s_-]+\frac{2}{3}[a(\frac{1}{2})(u_
{\langle l^{\pi^0}_{u_-}\rangle}+\pi^0_{\langle
l^{u_-}_{\pi^0}\rangle})\cr &+&a(\frac{\beta^2}{6})(u_{\langle
l^{\eta}_{u_-}\rangle}+\eta_{\langle
l^{u_-}_{\eta}\rangle})+a(\frac{\zeta^2}{3})(u_{\langle
l^{\eta'}_{u_-}\rangle}+\eta_{\langle
l^{u_-}_{\eta'}\rangle})+a(d_{\langle
l^{\pi^+}_{d_-}\rangle}+\pi^+_{\langle
l^{d_-}_{\pi^+}\rangle})+a\alpha^2(s_{\langle
l^{K^+}_{s_-}\rangle}+K^+_{\langle l^{s_-}_{K^+}\rangle})] \cr
&+&[1-a(\frac{9+\beta^2+2\zeta^2}{6}+\alpha^2)]d_+
+\frac{1}{3}[a(\frac{3+\beta^2+2\zeta^2}{6})d_-+au_-+a\alpha^2s_-]+\frac{2}{3}
[a(\frac{1}{2})(d_ {\langle l^{\pi^0}_{d_-}\rangle}+\pi^0_{\langle
l^{d_-}_{\pi^0}\rangle})\cr &+&a(\frac{\beta^2}{6})(d_{\langle
l^{\eta}_{d_-}\rangle}+\eta_{\langle
l^{d_-}_{\eta}\rangle})+a(\frac{\zeta^2}{3})(d_{\langle
l^{\eta'}_{d_-}\rangle}+\eta_{\langle
l^{d_-}_{\eta'}\rangle})+a(u_{\langle
l^{\pi^-}_{u_-}\rangle}+\pi^-_{\langle
l^{u_-}_{\pi^-}\rangle})+a\alpha^2(s_{\langle
l^{K^0}_{s_-}\rangle}+K^0_{\langle l^{s_-}_{K^0}\rangle})]\cr
&-&2[1-a(\frac{2\beta^2+\zeta^2}{3}+2\alpha^2)]s_+-\frac{1}{3}[\frac{2a}{3}(2
\beta^2+\zeta^2)s_-
+2a\alpha^2u_-+2a\alpha^2d_-]-\frac{2}{3}[\frac{4a}{3}\beta^2(s_{\langle
l^{\eta}_{s_-}\rangle}+\eta_{\langle l^{s_-}_{\eta}\rangle})\cr
&+&\frac{2a}{3}\zeta^2(s_{\langle
l^{\eta'}_{s_-}\rangle}+\eta'_{\langle
l^{s_-}_{\eta'}\rangle})+2a\alpha^2(u_{\langle
l^{K^-}_{u_-}\rangle}+K^-_{\langle
l^{u_-}_{K^-}\rangle})+2a\alpha^2(d_{\langle
l^{\bar{K}^0}_{d_-}\rangle}+\bar{K}^0_{\langle
l^{d_-}_{\bar{K}^0}\rangle})]\}\cdot(\textrm{i}\frac{1}{6} kR\cdot
e^{-\frac{1}{6}k^2R^2}),
\end{eqnarray}
\begin{eqnarray}
&&\widehat{\Lambda_{1520}\Lambda_{1116}}^*_{(\frac{3}{2})}(k)\cr
&=&\{[1-a(\frac{9+\beta^2+2\zeta^2}{6}+\alpha^2)]u
+a(\frac{1}{2})(u_{\langle l^{\pi^0}_{u}\rangle}+\pi^0_{\langle
l^{u}_{\pi^0}\rangle})+a(\frac{\beta^2}{6})(u_{\langle
l^{\eta}_{u}\rangle}+\eta_{\langle l^{u}_{\eta}\rangle})\cr
&+&a(\frac{\zeta^2}{3})(u_{\langle
l^{\eta'}_{u}\rangle}+\eta_{\langle
l^{u}_{\eta'}\rangle})+a(d_{\langle
l^{\pi^+}_{d}\rangle}+\pi^+_{\langle
l^{d}_{\pi^+}\rangle})+a\alpha^2(s_{\langle
l^{K^+}_{s}\rangle}+K^+_{\langle l^{s}_{K^+}\rangle})\cr
&+&[1-a(\frac{9+\beta^2+2\zeta^2}{6}+\alpha^2)]d
+a(\frac{1}{2})(d_{\langle l^{\pi^0}_{d}\rangle}+\pi^0_{\langle
l^{d}_{\pi^0}\rangle}) +a(\frac{\beta^2}{6})(d_{\langle
l^{\eta}_{d}\rangle}+\eta_{\langle l^{d}_{\eta}\rangle})\cr
&+&a(\frac{\zeta^2}{3})(d_{\langle
l^{\eta'}_{d}\rangle}+\eta_{\langle
l^{d}_{\eta'}\rangle})+a(u_{\langle
l^{\pi^-}_{u}\rangle}+\pi^-_{\langle
l^{u}_{\pi^-}\rangle})+a\alpha^2(s_{\langle
l^{K^0}_{s}\rangle}+K^0_{\langle l^{s}_{K^0}\rangle})\cr
&-&2[1-a(\frac{2\beta^2+\zeta^2}{3}+2\alpha^2)]s-\frac{4a}{3}\beta^2(s_
{\langle l^{\eta}_{s}\rangle}+\eta_{\langle l^{s}_{\eta}\rangle})
-\frac{2a}{3}\zeta^2(s_{\langle
l^{\eta'}_{s}\rangle}+\eta'_{\langle l^{s}_{\eta'}\rangle})\cr
&-&2a\alpha^2(u_{\langle l^{K^-}_{u}\rangle}+K^-_{\langle
l^{u}_{K^-}\rangle})-2a\alpha^2(d_{\langle
l^{\bar{K}^0}_{d}\rangle}+\bar{K}^0_{\langle
l^{d}_{\bar{K}^0}\rangle})\}\cdot(-\textrm{i}\frac{\sqrt{3}}{6}
\frac{1}{kR}\cdot e^{-\frac{1}{6}k^2R^2}),
\end{eqnarray}
\begin{eqnarray}
&&\widehat{\Lambda_{1520}\Lambda_{1116}}^*_{(\frac{1}{2})}(k)\cr
&=&\{[1-a(\frac{9+\beta^2+2\zeta^2}{6}+\alpha^2)]u
+a(\frac{1}{2})(u_{\langle l^{\pi^0}_{u}\rangle}+\pi^0_{\langle
l^{u}_{\pi^0}\rangle})+a(\frac{\beta^2}{6})(u_{\langle
l^{\eta}_{u}\rangle}+\eta_{\langle l^{u}_{\eta}\rangle})\cr
&+&a(\frac{\zeta^2}{3})(u_{\langle
l^{\eta'}_{u}\rangle}+\eta_{\langle
l^{u}_{\eta'}\rangle})+a(d_{\langle
l^{\pi^+}_{d}\rangle}+\pi^+_{\langle
l^{d}_{\pi^+}\rangle})+a\alpha^2(s_{\langle
l^{K^+}_{s}\rangle}+K^+_{\langle l^{s}_{K^+}\rangle})\cr
&+&[1-a(\frac{9+\beta^2+2\zeta^2}{6}+\alpha^2)]d
+a(\frac{1}{2})(d_{\langle l^{\pi^0}_{d}\rangle}+\pi^0_{\langle
l^{d}_{\pi^0}\rangle}) +a(\frac{\beta^2}{6})(d_{\langle
l^{\eta}_{d}\rangle}+\eta_{\langle l^{d}_{\eta}\rangle})\cr
&+&a(\frac{\zeta^2}{3})(d_{\langle
l^{\eta'}_{d}\rangle}+\eta_{\langle
l^{d}_{\eta'}\rangle})+a(u_{\langle
l^{\pi^-}_{u}\rangle}+\pi^-_{\langle
l^{u}_{\pi^-}\rangle})+a\alpha^2(s_{\langle
l^{K^0}_{s}\rangle}+K^0_{\langle l^{s}_{K^0}\rangle})\cr
&-&2[1-a(\frac{2\beta^2+\zeta^2}{3}+2\alpha^2)]s-\frac{4a}{3}\beta^2(s_
{\langle l^{\eta}_{s}\rangle}+\eta_{\langle l^{s}_{\eta}\rangle})
-\frac{2a}{3}\zeta^2(s_{\langle
l^{\eta'}_{s}\rangle}+\eta'_{\langle l^{s}_{\eta'}\rangle})\cr
&-&2a\alpha^2(u_{\langle l^{K^-}_{u}\rangle}+K^-_{\langle
l^{u}_{K^-}\rangle})-2a\alpha^2(d_{\langle
l^{\bar{K}^0}_{d}\rangle}+\bar{K}^0_{\langle
l^{d}_{\bar{K}^0}\rangle})\}\cdot(-\textrm{i}\frac{1}{6}
\frac{1}{kR}\cdot e^{-\frac{1}{6}k^2R^2})\;.
\end{eqnarray}
Using equations (34)-(37), we can calculate
$\mu^{total}_M(B_{i}B_f)$ and $\mu^{total}_M(B_{i}B_f)^*$
corresponding to (60), (61) respectively,
\begin{equation}
\mu^{total}_M(\Lambda_{1520}\Lambda_{1116})=\mu^{val}_M(\Lambda_{1520}\Lambda_
{1116})
+\mu^{sea}_M(\Lambda_{1520}\Lambda_{1116})+\mu^{orbit}_M(\Lambda_{1520}
\Lambda_{1116})\;,
\end{equation}
\begin{equation}
\mu^{val}_{(-\frac{1}{2})}(\Lambda_{1520}\Lambda_{1116})=(\mu_u+\mu_d
-2\mu_s)\cdot(\textrm{i}\frac{1} {6} kR\cdot
e^{-\frac{1}{6}k^2R^2})\;,
\end{equation}
\begin{equation}
\mu^{sea}_{(-\frac{1}{2})}(\Lambda_{1520}\Lambda_{1116})=-a[(\frac
{18+2\beta^2+4\zeta^2}{9}+\frac{1}{3}\alpha^2)(\mu_u+\mu_d)
-(\frac{16\beta^2+8\zeta^2}{9}+\frac{10}{3}\alpha^2)\mu_s]\cdot(\textrm{i}\frac{1}{6}
kR\cdot e^{-\frac{1}{6}k^2R^2})\;,
\end{equation}
\begin{eqnarray}
\mu^{orbit}_{(-\frac{1}{2})}(\Lambda_{1520}\Lambda_{1116})
&=&\frac{2}{3}a[\frac{1}{2}(\mu_u{\langle
l^{\pi^0}_{u_-}\rangle}+\mu_{\pi^0}{\langle
l^{u_-}_{\pi^0}\rangle})+\frac{\beta^2}{6}(\mu_u{\langle
l^{\eta}_{u_-}\rangle}+\mu_{\eta}{\langle
l^{u_-}_{\eta}\rangle})\cr &+&\frac{\zeta^2}{3}(\mu_{u}{\langle
l^{\eta'}_{u_-}\rangle}+\mu_{\eta}{\langle
l^{u_-}_{\eta'}\rangle})+(\mu_{d}{\langle
l^{\pi^+}_{d_-}\rangle}+\mu_{\pi^+}{\langle
l^{d_-}_{\pi^+}\rangle})+\alpha^2(\mu_s{\langle
l^{K^+}_{s_-}\rangle}+\mu_{K^+}{\langle l^{s_-}_{K^+}\rangle})\cr
&+&\frac{1}{2}(\mu_d{\langle
l^{\pi^0}_{d_-}\rangle}+\mu_{\pi^0}{\langle
l^{d_-}_{\pi^0}\rangle})+\frac{\beta^2}{6}(\mu_d{\langle
l^{\eta}_{d_-}\rangle}+\mu_{\eta}{\langle
l^{d_-}_{\eta}\rangle})\cr &+&\frac{\zeta^2}{3}(\mu_d{\langle
l^{\eta'}_{d_-}\rangle}+\mu_{\eta}{\langle
l^{d_-}_{\eta'}\rangle})+(\mu_u{\langle
l^{\pi^-}_{u_-}\rangle}+\mu_{\pi^-}{\langle
l^{u_-}_{\pi^-}\rangle})+\alpha^2(\mu_s{\langle
l^{K^0}_{s_-}\rangle}+\mu_{K^0}{\langle l^{s_-}_{K^0}\rangle})\cr
&-&\frac{4a}{3}\beta^2(\mu_s{\langle
l^{\eta}_{s_-}\rangle}+\mu_{\eta}{\langle l^{s_-}_{\eta}\rangle})
-\frac{2a}{3}\zeta^2(\mu_s{\langle
l^{\eta'}_{s_-}\rangle}+\mu_{\eta'}{\langle
l^{s_-}_{\eta'}\rangle})\cr &-&2a\alpha^2(\mu_u{\langle
l^{K^-}_{u_-}\rangle}+\mu_{K^-}{\langle
l^{u_-}_{K^-}\rangle})-2a\alpha^2(\mu_d{\langle
l^{\bar{K}^0}_{d_-}\rangle}+\mu_{\bar{K}^0}{\langle
l^{d_-}_{\bar{K}^0}\rangle})]\cdot(\textrm{i}\frac{1}{6} kR\cdot
e^{-\frac{1}{6}k^2R^2})\;,
\end{eqnarray}
\begin{equation}
\mu^{total}_M(\Lambda_{1520}\Lambda_{1116})^*=\mu^{val}_M(\Lambda_{1520}
\Lambda_{1116})^* +\mu^{sea}_M(\Lambda_{1520} \Lambda_{1116})^*
+\mu^{orbit}_M(\Lambda_{1520}\Lambda_{1116})^*\;,
\end{equation}
\begin{eqnarray}
\mu^{val}_{(\frac{3}{2})}(\Lambda_{1520}\Lambda_{1116})^*
&=&(\mu_u+\mu_d-2\mu_s)\cr &\cdot&(-\textrm{i}\frac{\sqrt{3}}{6}
\frac{1}{kR}\cdot e^{-\frac{1}{6}k^2R^2})\;,
\end{eqnarray}
\begin{eqnarray}
\mu^{sea}_{(\frac{3}{2})}(\Lambda_{1520}\Lambda_{1116})^*
&=&[-a(\frac{9+\beta^2+2\zeta^2}{6}+\alpha^2)](\mu_u+\mu_d)
-2[-a(\frac{2\beta^2+\zeta^2}{3}+2\alpha^2)]\mu_s\cr
&\cdot&(-\textrm{i}\frac{\sqrt{3}}{6} \frac{1}{kR}\cdot
e^{-\frac{1}{6}k^2R^2})\;,
\end{eqnarray}
\begin{eqnarray}
\mu^{orbit}_{(\frac{3}{2})}(\Lambda_{1520}\Lambda_{1116})^*
&=&a[\frac{1}{2}(\mu_u{\langle
l^{\pi^0}_{u}\rangle}+\mu_{\pi^0}{\langle
l^{u}_{\pi^0}\rangle})+\frac{\beta^2}{6}(\mu_u{\langle
l^{\eta}_{u}\rangle}+\mu_{\eta}{\langle l^{u}_{\eta}\rangle})\cr
&+&\frac{\zeta^2}{3}(\mu_{u}{\langle
l^{\eta'}_{u}\rangle}+\mu_{\eta}{\langle
l^{u}_{\eta'}\rangle})+(\mu_{d}{\langle
l^{\pi^+}_{d}\rangle}+\mu_{\pi^+}{\langle
l^{d}_{\pi^+}\rangle})+\alpha^2(\mu_s{\langle
l^{K^+}_{s}\rangle}+\mu_{K^+}{\langle l^{s}_{K^+}\rangle})\cr
&+&\frac{1}{2}(\mu_d{\langle
l^{\pi^0}_{d}\rangle}+\mu_{\pi^0}{\langle
l^{d}_{\pi^0}\rangle})+\frac{\beta^2}{6}(\mu_d{\langle
l^{\eta}_{d}\rangle}+\mu_{\eta}{\langle l^{d}_{\eta}\rangle})\cr
&+&\frac{\zeta^2}{3}(\mu_d{\langle
l^{\eta'}_{d}\rangle}+\mu_{\eta}{\langle
l^{d}_{\eta'}\rangle})+(\mu_u{\langle
l^{\pi^-}_{u}\rangle}+\mu_{\pi^-}{\langle
l^{u}_{\pi^-}\rangle})+\alpha^2(\mu_s{\langle
l^{K^0}_{s}\rangle}+\mu_{K^0}{\langle l^{s}_{K^0}\rangle})\cr
&-&\frac{4a}{3}\beta^2(\mu_s{\langle
l^{\eta}_{s}\rangle}+\mu_{\eta}{\langle l^{s}_{\eta}\rangle})
-\frac{2a}{3}\zeta^2(\mu_s{\langle
l^{\eta'}_{s}\rangle}+\mu_{\eta'}{\langle
l^{s}_{\eta'}\rangle})\cr &-&2a\alpha^2(\mu_u{\langle
l^{K^-}_{u}\rangle}+\mu_{K^-}{\langle
l^{u}_{K^-}\rangle})-2a\alpha^2(\mu_d{\langle
l^{\bar{K}^0}_{d}\rangle}+\mu_{\bar{K}^0}{\langle
l^{d}_{\bar{K}^0}\rangle})]\cdot(-\textrm{i}\frac{\sqrt{3}}{6}
\frac{1}{kR}\cdot e^{-\frac{1}{6}k^2R^2})\;,
\end{eqnarray}
\begin{eqnarray}
\mu^{val}_{(\frac{1}{2})}(\Lambda_{1520}\Lambda_{1116})^*
&=&(\mu_u+\mu_d-2\mu_s)\cr &\cdot&(-\textrm{i}\frac{1}{6}
\frac{1}{kR}\cdot e^{-\frac{1}{6}k^2R^2})\;,
\end{eqnarray}
\begin{eqnarray}
\mu^{sea}_{(\frac{1}{2})}(\Lambda_{1520}\Lambda_{1116})^*
&=&[-a(\frac{9+\beta^2+2\zeta^2}{6}+\alpha^2)](\mu_u+\mu_d)
-2[-a(\frac{2\beta^2+\zeta^2}{3}+2\alpha^2)]\mu_s\cr
&\cdot&(-\textrm{i}\frac{1}{6} \frac{1}{kR}\cdot
e^{-\frac{1}{6}k^2R^2})\;,
\end{eqnarray}
\begin{eqnarray}
\mu^{orbit}_{(\frac{1}{2})}(\Lambda_{1520}\Lambda_{1116})^*
&=&a[\frac{1}{2}(\mu_u{\langle
l^{\pi^0}_{u}\rangle}+\mu_{\pi^0}{\langle
l^{u}_{\pi^0}\rangle})+\frac{\beta^2}{6}(\mu_u{\langle
l^{\eta}_{u}\rangle}+\mu_{\eta}{\langle l^{u}_{\eta}\rangle})\cr
&+&\frac{\zeta^2}{3}(\mu_{u}{\langle
l^{\eta'}_{u}\rangle}+\mu_{\eta}{\langle
l^{u}_{\eta'}\rangle})+(\mu_{d}{\langle
l^{\pi^+}_{d}\rangle}+\mu_{\pi^+}{\langle
l^{d}_{\pi^+}\rangle})+\alpha^2(\mu_s{\langle
l^{K^+}_{s}\rangle}+\mu_{K^+}{\langle l^{s}_{K^+}\rangle})\cr
&+&\frac{1}{2}(\mu_d{\langle
l^{\pi^0}_{d}\rangle}+\mu_{\pi^0}{\langle
l^{d}_{\pi^0}\rangle})+\frac{\beta^2}{6}(\mu_d{\langle
l^{\eta}_{d}\rangle}+\mu_{\eta}{\langle l^{d}_{\eta}\rangle})\cr
&+&\frac{\zeta^2}{3}(\mu_d{\langle
l^{\eta'}_{d}\rangle}+\mu_{\eta}{\langle
l^{d}_{\eta'}\rangle})+(\mu_u{\langle
l^{\pi^-}_{u}\rangle}+\mu_{\pi^-}{\langle
l^{u}_{\pi^-}\rangle})+\alpha^2(\mu_s{\langle
l^{K^0}_{s}\rangle}+\mu_{K^0}{\langle l^{s}_{K^0}\rangle})\cr
&-&\frac{4a}{3}\beta^2(\mu_s{\langle
l^{\eta}_{s}\rangle}+\mu_{\eta}{\langle l^{s}_{\eta}\rangle})
-\frac{2a}{3}\zeta^2(\mu_s{\langle
l^{\eta'}_{s}\rangle}+\mu_{\eta'}{\langle
l^{s}_{\eta'}\rangle})\cr &-&2a\alpha^2(\mu_u{\langle
l^{K^-}_{u}\rangle}+\mu_{K^-}{\langle
l^{u}_{K^-}\rangle})-2a\alpha^2(\mu_d{\langle
l^{\bar{K}^0}_{d}\rangle}+\mu_{\bar{K}^0}{\langle
l^{d}_{\bar{K}^0}\rangle})]\cdot(-\textrm{i}\frac{1}{6}
\frac{1}{kR}\cdot e^{-\frac{1}{6}k^2R^2})\;.
\end{eqnarray}
Considering the relationship between $\mu_M(B_{i}B_f)$,
$\mu_M(B_{i}B_f)^*$ and $A_M^{M1}$, $A_M^{E1}$ for
$\Lambda(1520)\rightarrow\Lambda(1116)+\gamma$, we have
\begin{equation}
A_{\frac{3}{2}}^{total}(\Lambda_{1520}\rightarrow\Lambda_{1116}+\gamma)
=A_{\frac{3}{2}}^{E1,val}(\Lambda_{1520}\rightarrow\Lambda_{1116}+\gamma)
+A_{\frac{3}{2}}^{E1,sea}(\Lambda_{1520}\rightarrow\Lambda_{1116}+\gamma)
+A_{\frac{3}{2}}^{E1,orbit}(\Lambda_{1520}\rightarrow\Lambda_{1116}+\gamma),
\end{equation}
$(A_{\frac{3}{2}}^{M1}(\Lambda_{1520}\rightarrow\Lambda_{1116}+\gamma)=0),$
where,
\begin{equation}
A_{\frac{3}{2}}^{E1,val}(\Lambda_{1520}\rightarrow\Lambda_{1116}+\gamma)
=2\sqrt{{\pi}{k}}\cdot[\mu^{val}_{\frac{3}{2}}(\Lambda_{1520}\Lambda_{1116})
^*]
\end{equation}
\begin{equation}
A_{\frac{3}{2}}^{E1,sea}(\Lambda_{1520}\rightarrow\Lambda_{1116}+\gamma)
=2\sqrt{{\pi}{k}}\cdot[\mu^{sea}_{\frac{3}{2}}(\Lambda_{1520}\Lambda_{1116})
^*]
\end{equation}
\begin{equation}
A_{\frac{3}{2}}^{E1,orbit}(\Lambda_{1520}\rightarrow\Lambda_{1116}+\gamma)
=2\sqrt{{\pi}{k}}\cdot[\mu^{orbit}_{\frac{3}{2}}(\Lambda_{1520}\Lambda_{1116})
^*]
\end{equation}
\begin{eqnarray}
A_{\frac{1}{2}}^{total}(\Lambda_{1520}\rightarrow\Lambda_{1116}+\gamma)
&=&A_{\frac{1}{2}}^{M1,val}(\Lambda_{1520}\rightarrow\Lambda_{1116}+\gamma)
+A_{\frac{1}{2}}^{M1,sea}(\Lambda_{1520}\rightarrow\Lambda_{1116}+\gamma)\cr
&+&A_{\frac{1}{2}}^{M1,orbit}(\Lambda_{1520}\rightarrow\Lambda_{1116}+\gamma)
+A_{\frac{1}{2}}^{E1,val}(\Lambda_{1520}\rightarrow\Lambda_{1116}+\gamma)\cr
&+&A_{\frac{1}{2}}^{E1,sea}(\Lambda_{1520}\rightarrow\Lambda_{1116}+\gamma)
+A_{\frac{1}{2}}^{E1,orbit}(\Lambda_{1520}\rightarrow\Lambda_{1116}+\gamma)
\end{eqnarray}
where,
\begin{equation}
A_{\frac{1}{2}}^{M1,val}(\Lambda_{1520}\rightarrow\Lambda_{1116}+\gamma)
=2\sqrt{{\pi}{k}}\cdot[\mu^{val}_{-\frac{1}{2}}(\Lambda_{1520}\Lambda_{1116})]
\end{equation}
\begin{equation}
A_{\frac{1}{2}}^{M1,sea}(\Lambda_{1520}\rightarrow\Lambda_{1116}+\gamma)
=2\sqrt{{\pi}{k}}\cdot[\mu^{sea}_{-\frac{1}{2}}(\Lambda_{1520}\Lambda_{1116})]
\end{equation}
\begin{equation}
A_{\frac{1}{2}}^{M1,orbit}(\Lambda_{1520}\rightarrow\Lambda_{1116}+\gamma)
=2\sqrt{{\pi}{k}}\cdot[\mu^{orbit}_{-\frac{1}{2}}(\Lambda_{1520}\Lambda_
{1116})]
\end{equation}
\begin{equation}
A_{\frac{1}{2}}^{E1,val}(\Lambda_{1520}\rightarrow\Lambda_{1116}+\gamma)
=2\sqrt{{\pi}{k}}\cdot[\mu^{val}_{\frac{1}{2}}(\Lambda_{1520}\Lambda_{1116})
^*]
\end{equation}
\begin{equation}
A_{\frac{1}{2}}^{E1,sea}(\Lambda_{1520}\rightarrow\Lambda_{1116}+\gamma)
=2\sqrt{{\pi}{k}}\cdot[\mu^{sea}_{\frac{1}{2}}(\Lambda_{1520}\Lambda_{1116})
^*]
\end{equation}
\begin{equation}
A_{\frac{1}{2}}^{E1,orbit}(\Lambda_{1520}\rightarrow\Lambda_{1116}+\gamma)
=2\sqrt{{\pi}{k}}\cdot[\mu^{orbit}_{\frac{1}{2}}(\Lambda_{1520}\Lambda_{1116})
^*]
\end{equation}
besides,
\begin{equation}
A_{\frac{3}{2}}^{total}(\Lambda_{1520}\rightarrow\Lambda_{1116}+\gamma)
=2\sqrt{{\pi}{k}}\cdot[\mu^{total}_{\frac{3}{2}}(\Lambda_{1520}\Lambda_{1116})
^*]
\end{equation}
\begin{equation}
A_{\frac{1}{2}}^{total}(\Lambda_{1520}\rightarrow\Lambda_{1116}+\gamma)
=2\sqrt{{\pi}{k}}\cdot[\mu^{total}_{-\frac{1}{2}}(\Lambda_{1520}\Lambda_
{1116})
+\mu^{total}_{\frac{1}{2}}(\Lambda_{1520}\Lambda_{1116})^*]
\end{equation}

\section{RESULTS AND CONCLUSIONS}
\label{sec-summary}

We collect the input parameters $a$, $\alpha$, $\beta$, $\zeta$,
$R$ (the harmonic-oscillator radius parameter) and the masses of
the quarks and $GBs$ in Table I. The parameters $a$, $\alpha$,
$\beta$ and $\zeta$ are fixed by fitting the octet baryon magnetic
moments \cite{XQM4,XQM6,XQM7,MM2}. We have used the commonly used values
in hadron spectroscopy for $R$ and constitutent quark masses
\cite{WF,RT,HS}. And we use physical masses for $GBs$
\cite{XQM5,MM2}.
\vspace{1cm}
\begin{center}
\setlength{\tabcolsep}{3mm}
\begin{tabular}{c|c|c|c|c|c|c|c}
\hline {Input}&{$a$}&{$\alpha$}
&{$\beta$}&{$\zeta$}&{$R(GeV^{-1})$}&{$M_{u,d}(MeV)$}&{$M_s(MeV)$}\\\hline
Value&{0.1}&{0.4} &{0.4}&{-0.4}&{2.45}&{350}&{500}\\\hline
\end{tabular}
\end{center}
\begin{center}
\begin{tabular}{c}{Table I. The values of various inputs used in our
calculation.}
\end{tabular}
\end{center}

With the above parameters, the octet magnetic moments in the chiral quark model
are listed in Table II. Numerically speaking, the sea quark and orbital
contributions to the octet baryon magnetic moments are both quite large in magnitude
except for $\Xi^-$ and $\Lambda$. However, their contributions cancel each other
to a large extent. The sum of the residual sea and the naive valence quark
contribution agrees with experimental value quite well as can be seen from Table II.
\vspace{1cm}
\begin{center}
\begin{tabular}{c|c|c|c|c|c}
\hline Octet&Exp.&\multicolumn{4}{c}{$\chi QM$}\\\cline{3-6}
Baryons&\cite{DATA}&$\mu^{val}$& $\mu^{sea}$&$\mu^{orbit}$&
$\mu^{total}$\\\hline
$p$&2.794&2.680&-0.455&0.569&2.794\\
$n$&-1.913&-1.787&0.259&-0.537&-2.065\\
$\Sigma^{+}$&2.458&2.591&-0.426&0.466&2.631\\
$\Sigma^{-}$&-1.160&-0.983&0.146&-0.419&-1.256\\
$\Xi^{0}$&-1.250&-1.429&0.142&-0.125&-1.412\\
$\Xi^{-}$&0.651&-0.536&-0.001&0.097&-0.440\\
$\Lambda$&-0.613&-0.625&0.029&-0.007&-0.603\\
$\Sigma^{0}\Lambda$&1.61&1.547&-0.248&0.383&1.682\\
 \hline
\end{tabular}
\end{center}
\begin{center}
\begin{tabular}{c}{Table II. The octet magnetic
moments in units of $\mu_N$}
\end{tabular}
\end{center}

With the same parameters, we present the results of decuplet to octet
transition magnetic moments in Table III.
\vspace{1cm}
\begin{center}
\begin{tabular}{c|c|c|c|c|c}
\hline &\multicolumn{5}{c}{$\chi QM$}\\\hline $B_{10}\rightarrow
B_8+\gamma$&$\mu_{B_{10}B_8}^{val}$&
$\mu_{B_{10}B_8}^{sea}$&$\mu_{B_{10}B_8}^{orbit}$&
$\mu_{B_{10}B_8}^{total}$&$\mu_{B_{10}B_8}^{total}(k)$\\\hline
$\Delta^+\rightarrow
p+\gamma$&2.527&-0.404&0.626&2.749&2.572\\$\Delta^0\rightarrow
n+\gamma$&2.527&-0.404&0.626&2.749&2.572\\\hline
$\Sigma^{*,+}\rightarrow \Sigma^++\gamma$&-2.274&0.321&-0.334&-2.287&-2.215\\
$\Sigma^{*,0}\rightarrow \Sigma^0+\gamma$&-1.011&0.119&-0.021&-0.913&-0.884\\
$\Sigma^{*,0}\rightarrow \Lambda+\gamma$&2.188&-0.350&0.542&2.380&2.245\\
$\Sigma^{*,-}\rightarrow
\Sigma^-+\gamma$&0.253&-0.083&0.292&0.462&0.447\\\hline
$\Xi^{*,0}\rightarrow \Xi^0+\gamma$&-2.274&0.321&-0.334&-2.287&-2.215\\
$\Xi^{*,-}\rightarrow \Xi^-+\gamma$&0.253&-0.083&0.292&0.462&0.447\\
\hline
\end{tabular}
\end{center}
\begin{center}
\begin{tabular}{c}{Table III. The decuplet to octet transition magnetic
moments in units of $\mu_N$}
\end{tabular}
\end{center}

Since the configuration mixing effect can not be ignored for the
low-lying excited hyperons, we need include the configuration mixing
terms in the hyperon wave functions. For the $\Lambda(1520)$ and
$\Lambda(1405)$, we take the following wave functions from
the previous analysis \cite{QM2,QM4}:
\begin{equation}
| \Lambda(1520),\frac{3}{2}^- \rangle= 0.91| \Lambda\,70,\,
^21,1,1,\frac{3}{2}^-\rangle-0.40
|\Lambda\,70,\,^28,1,1,\frac{3}{2}^- \rangle+0.01
|\Lambda\,70,\,^48,1,1,\frac{3}{2}^- \rangle,
\end{equation}
\begin{equation}
| \Lambda(1405),\frac{1}{2}^- \rangle= 0.90
|\Lambda\,70,\,^21,1,1,\frac{1}{2}^- \rangle-0.43
|\Lambda\,70,\,^28,1,1,\frac{1}{2}^- \rangle-0.06
|\Lambda\,70,\,^48,1,1,\frac{1}{2}^- \rangle,
\end{equation}
where the baryon $SU(6)\otimes O(3)$ wave function is defined as
$|N_6,\,^{2S+1}N_3,N,L,J^P \rangle$\cite{WF}. Here $N_6$ and $N_3$
are the $SU(6)$ and $SU(3)$ multiplicity respectively, while $S$,
$N$, $L$, $J$ and $P$ are the total spin, the radial quantum
number, total orbital angular momentum, total angular momentum and
parity.

We present the helicity amplitudes of various radiative decays in the chiral quark model
in Table IV. These helicity amplitudes are decomposed into valence quark contribution,
sea contribution and orbital contribution respectively.
\vspace{1cm}
\begin{center}
\begin{tabular}{c|c|c|c|c|c|c|c|c|c|c|c|c|c|c}
\hline &\multicolumn{14}{c}{$\chi
QM$}\\\cline{2-15}$B_{i}\rightarrow B_f+\gamma$
&\multicolumn{3}{c|}{$A_{\frac{3}{2}}^{M1}$}
&\multicolumn{3}{c|}{$A_{\frac{3}{2}}^{E1}$}&$A_{\frac{3}{2}}$
&\multicolumn{3}{c|}{$A_{\frac{1}{2}}^{M1}$}
&\multicolumn{3}{c|}{$A_{\frac{1}{2}}^{E1}$}&$A_{\frac{1}{2}}$\\\cline{2-15}
&$val$&$sea$&$orbit$&$val$&$sea$&$orbit$&$total$&$val$&$sea$&$orbit$&
$val$&$sea$&$orbit$&$total$\\\hline $\Delta^+\rightarrow
p+\gamma$&-0.168&0.027&-0.042&0&0&0&-0.183&-0.097&0.016&-0.024&0&0&0&-0.105\\$\Delta^0\rightarrow
n+\gamma$&-0.168&0.027&-0.042&0&0&0&-0.183&-0.097&0.016&-0.024&0&0&0&-0.105\\\hline
$\Sigma^{*,+}\rightarrow \Sigma^++\gamma$&0.130&-0.018&0.019&0&0&0&0.131&0.075&-0.010&0.011&0&0&0&0.076\\
$\Sigma^{*,0}\rightarrow \Sigma^0+\gamma$&0.058&-0.007&0.001&0&0&0&0.052&0.033&-0.004&0.001&0&0&0&0.030\\
$\Sigma^{*,0}\rightarrow \Lambda+\gamma$&-0.142&0.023&-0.035&0&0&0&-0.154&-0.082&0.013&-0.020&0&0&0&-0.089\\
$\Sigma^{*,-}\rightarrow
\Sigma^-+\gamma$&-0.014&0.005&-0.017&0&0&0&-0.026&-0.008&0.003&-0.010&0&0&0&-0.015\\\hline
$\Xi^{*,0}\rightarrow \Xi^0+\gamma$&0.136&-0.019&0.020&0&0&0&0.137&0.078&-0.011&0.012&0&0&0&0.079\\
$\Xi^{*,-}\rightarrow \Xi^-+\gamma$&-0.015&0.005&-0.017&0&0&0&-0.027&-0.009&0.003&-0.010&0&0&0&-0.016\\
\hline $|\Lambda\,70,\,^21,1,1,\frac{3}{2}^- \rangle
$&&&&&&&&&&&&&&\\
$\rightarrow\Lambda+\gamma$&0&0&0&-0.061&0.006&-0.002&-0.057&0.026&-0.003&-0.001&-0.035&0.004&-0.001&-0.010\\
$|\Lambda\,70,\,^28,1,1,\frac{3}{2}^-
\rangle$&&&&&&&&&&&&&&\\
$\rightarrow\Lambda+\gamma$&0&0&0&0.061&-0.006&0.002&0.057&-0.004&-0.002&0&0.035&-0.003&0.001&0.027\\
$|\Lambda\,70,\,^48,1,1,\frac{3}{2}^- \rangle$&&&&&&&&&&&&&&\\
$\rightarrow\Lambda+\gamma$&0.018&-0.004&0&0&0&0&0.014&0.004&-0.001&0&0&0&0&0.003\\\hline
$|\Lambda\,70,\,^21,1,1,\frac{3}{2}^- \rangle$&&&&&&&&&&&&&&\\
$\rightarrow\Sigma^0+\gamma$&0&0&0&-0.151&0.026&-0.056&-0.181&0.044&-0.007&0.011&-0.087&0.015&-0.032&-0.056\\
$|\Lambda\,70,\,^28,1,1,\frac{3}{2}^-
\rangle$&&&&&&&&&&&&&&\\
$\rightarrow\Sigma^0+\gamma$&0&0&0&-0.151&0.026&-0.056&-0.181&0.015&-0.002&0.004&-0.087&0.015&-0.032&-0.087\\
$|\Lambda\,70,\,^48,1,1,\frac{3}{2}^-
\rangle$&&&&&&&&&&&&&&\\
$\rightarrow\Sigma^0+\gamma$&0&0&0&-0.024&0.004&-0.006&-0.026&0&0&0&-0.005&0.001&-0.001&-0.005\\\hline
$|\Lambda\,70,\,^21,1,1,\frac{1}{2}^- \rangle$&&&&&&&&&&&&&&\\
$\rightarrow\Lambda+\gamma$&0&0&0&0&0&0&0&-0.012&0.001&0&-0.061&0.006&-0.002&-0.068\\
$|\Lambda\,70,\,^28,1,1,\frac{1}{2}^-
\rangle$&&&&&&&&&&&&&&\\
$\rightarrow\Lambda+\gamma$&0&0&0&0&0&0&0&0.002&0.001&0&0.061&-0.006&0.002&0.060\\
$|\Lambda\,70,\,^48,1,1,\frac{1}{2}^-
\rangle$&&&&&&&&&&&&&&\\
$\rightarrow\Lambda+\gamma$&0&0&0&0&0&0&0&0.010&-0.002&0&0&0&0&0.008\\\hline
$|\Lambda\,70,\,^21,1,1,\frac{1}{2}^-
\rangle$&&&&&&&&&&&&&&\\
$\rightarrow\Sigma^0+\gamma$&0&0&0&0&0&0&0&-0.018&0.003&-0.004&-0.157&0.027&-0.058&-0.207\\
$|\Lambda\,70,\,^28,1,1,\frac{1}{2}^-
\rangle$&&&&&&&&&&&&&&\\
$\rightarrow\Sigma^0+\gamma$&0&0&0&0&0&0&0&-0.006&0.001&-0.001&-0.157&0.027&-0.058&-0.194\\
$|\Lambda\,70,\,^48,1,1,\frac{1}{2}^-
\rangle$&&&&&&&&&&&&&&\\
$\rightarrow\Sigma^0+\gamma$&0&0&0&0&0&0&0&-0.012&0.002&-0.003&0&0&0&-0.013\\\hline
\end{tabular}
\end{center}
\begin{center}
{Table IV. The helicity amplitudes for the radiative transitions
(in GeV$^{-\frac{1}{2}}$)}
\end{center}

In Table V, we present the radiative decay widths of decuplet
baryons and excited $\Lambda$ hyperons in the chiral quark model.
Moreover, we collect all the available calculations of these
processes in literature and experimental data in Table V.
\vspace{1cm}
\begin{center}
\begin{tabular}{c|c|c|c|c|c|c|c}
\hline  $B_{i}\rightarrow B_f+\gamma$&{$\chi
QM$}&$NRQM$\cite{QM3,QM4}&$RCQM$\cite{RCQM}&$MIT\;Bag$\cite{QM3}&$Chiral\;
Bag$\cite{CBM}&$Skyrme$ \cite{SKM}&$Solition$\cite{SOM}\\\hline
$\Delta^+\rightarrow
p+\gamma$&363&360&&&&309-348&\\$\Delta^0\rightarrow
n+\gamma$&363&360&&&&309-348&\\\hline
$\Sigma^{*,+}\rightarrow \Sigma^++\gamma$&100&&&&&&\\
$\Sigma^{*,0}\rightarrow \Sigma^0+\gamma$&16&22&23&15&&7.7-16&19,11\\
$\Sigma^{*,0}\rightarrow \Lambda+\gamma$&241&273&267&152&&157-209&243,170\\
$\Sigma^{*,-}\rightarrow \Sigma^-+\gamma$&4.1&&&&&&\\\hline
$\Xi^{*,0}\rightarrow \Xi^0+\gamma$&133&&&&&&\\
$\Xi^{*,-}\rightarrow \Xi^-+\gamma$&5.4&&&&&&\\\hline
$\Lambda_{1405}\rightarrow \Lambda_{1116}+\gamma$&131&200&118&60,17&75&&44,40\\
$\Lambda_{1405}\rightarrow
\Sigma^0_{1193}+\gamma$&109&72&46&18,2.7&1.9&&13,17\\\hline
$\Lambda_{1520}\rightarrow \Lambda_{1116}+\gamma$&85&156&215&46&32&&\\
$\Lambda_{1520}\rightarrow
\Sigma^0_{1193}+\gamma$&94&55&293&17&51&&\\
\hline
\end{tabular}
\end{center}
\begin{center}
\begin{tabular}{c|c|c|c|c|c}
\hline $Algebaic\; model$\cite{AM}&$HB\chi PT$ \cite{HBPT}&$1/N_c
\;expansion$\cite{NM}&$Lattice$\cite{LA}&Previous Exp.&JLAB
Exp.\cite{CLAS}\\\hline
343.7&670-790&&430&640-720\cite{EX1}&\\ 343.7&670-790&&430&640-720&\\\hline &&&100&&\\
33.9&1.4-36&$24.9\pm4.1$&17&$<1750$\cite{EX2}&\\ 221.3&290-470&$298\pm25$&&$<2000$\cite{EX2}&$479\pm120^{+81}_{-100}$\\
&&&3.3&&\\\hline &&&129&&\\&&&3.8&&\\\hline
116.9&&&&$27\pm8$\cite{EX3}&\\155.7&&&&$10\pm4$\cite{EX3},$23\pm7$\cite{EX3}&\\\hline
85.1&&&&$33\pm11$\cite{EX4},$134\pm23$\cite{EX5}&$167\pm43^{+26}_{-12}$\\180.4&&&&$47\pm17$\cite{EX4}&\\\hline
\end{tabular}
\end{center}
\begin{center}
\begin{tabular}{l} Table V. The radiative widths (in keV) of the theoretical
predictions and experimental values for the radiative
transitions.
\end{tabular}
\end{center}

With these tables, we compare the contribution of the quark sea
with that of the valence quarks, which gives a modification of the
NRQM. For example in Table III, the orbital part contributes with
the same sign as the sum of the valence quark contribution, while
the sea part contributes with the opposite sign. Especially for
these two decays: $\Delta^+\rightarrow p+\gamma$ and
$\Sigma^{*,0}\rightarrow \Lambda+\gamma$, the experimental values
are higher than the valence contribution. We find that the total
contribution from the quark sea is positive and increases the
valence contribution by 10\%. Another important observation is the
large cancellation between the orbital and sea quark contribution.

For the hyperons in Table IV, the amplitudes of the
magnetic-dipole transitions and electric-dipole transitions from
the quark sea are not more than 10\% of the valence contribution
in most cases. But for the processes
$|\Lambda\,70,\,^21,1,1,\frac{1}{2}^-
\rangle\rightarrow\Sigma^0+\gamma$ and
$|\Lambda\,70,\,^21,1,1,\frac{3}{2}^-
\rangle\rightarrow\Sigma^0+\gamma$, the quark sea contribution is
significant, which is around 20\% of the valence quark
contribution. For these hyperons, configuration mixing effects are
also important. The radiative decays from the chiral quark model
roughly agrees with recent JLAB measurement. Hopefully this model
can be further extended to calculate other interesting
observables.

\section*{ACKNOWLEDGMENT}
This project was supported by the National Natural Science
Foundation of China under Grants 10375003 and 10421503, Ministry
of Education of China, FANEDD, Key Grant Project of Chinese
Ministry of Education (NO 305001) and SRF for ROCS, SEM.


\end{document}